\documentclass[prb,twocolumn,showpacs,preprintnumbers,amsmath,amssymb]{revtex4}
\usepackage{graphicx}% Include figure files
\usepackage{dcolumn}% Align table columns on decimal point
\usepackage{bm}% bold math
\usepackage{psfig}
\newcommand \be{\begin{eqnarray}}
\newcommand \ee{\end{eqnarray}}
\begin{document}
%\twocolumn[\hsize\textwidth\columnwidth\hsize
%           \csname @twocolumnfalse\endcsname
\title{Surface energy and magneto-capacitance of superconductors under electric field bias} 
\author{K. Morawetz$^{1,2}$, P. Lipavsk\'y$^{3,4}$, J. Kol\'a\v cek$^4$ and E. H. Brandt$^5$
% and M. Schreiber$^2$
}
\affiliation{$^1$Forschungszentrum Rossendorf, PF 51 01 19, 01314 Dresden, Germany}
%$^1$Institute of Physics, Chemnitz University of Technology, 
%09107 Chemnitz, Germany}
\affiliation{
$^2$Max-Planck-Institute for the Physics of Complex
Systems, Noethnitzer Str. 38, 01187 Dresden, Germany}
\affiliation{
$^3$Faculty of Mathematics and Physics, Charles University, 
Ke Karlovu 3, 12116 Prague 2, Czech Republic}
\affiliation{
$^4$Institute of Physics, Academy of Sciences, 
Cukrovarnick\'a 10, 16253 Prague 6, Czech Republic
}
\affiliation{
$^5$Max-Planck-Institute for Metals Research,
         D-70506 Stuttgart, Germany}
\begin{abstract}
A superconducting layer exposed to a perpendicular electric field and a parallel magnetic 
field is considered within the Ginzburg-Landau (GL) approach. The GL equation is 
solved near the surface and the surface energy is calculated. The  
nucleation critical field of superconducting state at the surface depends on the magnetic and electric 
fields. Special consideration is paid to the induced magnetic-field effect caused by diamagnetic surface 
currents. The latter effect is strongly dependent on the thickness of the sample. The effective inverse capacitance determines the effective penetration depth. It is found that the capacitance exhibits a jump at the surface critical field. An experiment is suggested for determining the change 
in the effective capacitance of the layer.
\end{abstract}
\pacs{
%03.65.Nk %      Scattering theory
%,21.45.+v %     Few-body systems
%,72.10.Fk %     Scattering by point defects, dislocations, surfaces, and other imperfections (including Kondo effect)
%,03.65.Ge %     Solutions of wave equations: bound states
%,34.80.Pa %     Coherence and correlation in electron scattering
%,34.10.+x %     General theories and models of atomic and molecular collisions and interactions (including statistical theories, transition state, stochastic and trajectory models, etc.)
%,68.65.Hb %Quantum dots
%,73.22.-f %     Electronic structure of nanoscale materials: clusters, nanoparticles, nanotubes, and nanocrystals
%,79.20.Rf %    Atomic, molecular, and ion beam impact and interactions with surfaces
%, 61.14.Dc %    Theories of diffraction and scattering
%,61.46.+w %     Nanoscale materials: clusters, nanoparticles, nanotubes, and nanocrystals 
%71.45.Gm, %      Exchange, correlation, dielectric and magnetic response functions, plasmons
%78.20.-e, %      Optical properties of bulk materials and thin films 
%78.47.+p, %      Time-resolved optical spectroscopies and other ultrafast optical measurements in condensed matter 
%42.65.Re, %     Ultrafast processes; optical pulse generation and pulse compression
%82.53.Mj %     Femtosecond probing of semiconductor nanostructures 
74.25.Op, % Mixed states, critical fields, and surface sheaths
74.25.Nf, %Response to electromagnetic fields (nuclear magnetic resonance, surface impedance, etc.)
74.20.De, %     Phenomenological theories (two-fluid,
          %                                  Ginzburg-Landau, etc.)
%74.25.Ld, %     Mechanical and acoustical properties, elasticity,
          %                              and ultrasonic attenuation
%74.25.Qt, %     Vortex lattices, flux pinning, flux creep
%74.81.-g%     Inhomogeneous superconductors
          %                            and superconducting systems
85.25.-j %Superconducting devices
}
\maketitle
%    \vskip2pc]

\section{Introduction}
The gate voltage can be used to change the carrier density of
superconducting surfaces and therefore the critical temperature 
in the same manner as in field effect 
semiconductor devices. This effect
has been investigated for more than 40 years \cite{GS60} and 
has continuously gained experimental interest,\cite{MBCAHM94,SOKLCCG97,WHKG99,ATM03,MGT03} see also the overview.\cite{KS98}
The critical temperature of thin superconducting layers 
can be controlled in this way by an electric field applied 
perpendicular to the layer.\cite{GS60,XDWKLV92,FreyMannhart95,ATM03,MGT03} 
High-$T_c$ superconductors are characterized by a low density of carriers such that this field effect is expected to be higher; this has caused a wide experimental activity .\cite{Ma92,MSBG93,MBCAHM94,KSKTB97,SOKLCCG97,ABHS98} Such field effect devices may be made even from organic and macromolecular films.\cite{CBBA06} 

According to the Anderson theorem the bias voltage can change the critical temperature only indirectly via the electric field dependence of the material parameters.\cite{Ric69a,BKS93,UGD02,FMBW95,SCLGPV04,CVG94,WHKG99} 
The influence of the electric field on the pairing mechanism is therefore to be expected in the density of states for very pure and thin low-dimensional structures \cite{gM96,MOR94,Mtc01} analogously to the
formation of sidebands in the density of states due to high fields.\cite{BJ92} Capacitance measurements on surfaces of high-temperature superconductors have revealed a so far unknown mechanism for electric-field penetration. \cite{JT93}   

On the other hand, magneto-capacitance techniques are used frequently to measure the influence of the magnetic field \cite{MHA03}, e.g. to test spin-dependent electrochemical potentials.  Starting from the early reports on an increase of the capacitance \cite{SSS66} for superconducting tunneling junctions \cite{PSS69}, the residual surface resistance of superconducting resonators is still under discussion 
%\cite{H71}
\cite{HAE87,B00jz,B08} since it becomes important for the question how short the electron bunches can be in free electron lasers before the generated wake fields disable the superconducting cavities. \cite{B00jz} Different mechanisms for such electron losses have been discussed in \cite{B08}. In this respect it is important to know the explicit dependence on the magnetic field and the voltage bias.

In this paper we investigate the magneto-capacitance in dependence on the magnetic field and the external bias by the electric field. 
We focus on magnetic fields around the  
surface critical field $B_{\rm c3}$ 
since we expect that the external bias, which affects 
only the surface, 
has a relatively large effect on the surface superconductivity.
Most experimental activities are concentrated on the change of the surface critical magnetic field with temperature.\cite{TJ64,HK69,OF69,IMOT71,BB74,B84}
In our study we suggest to consider these measurements under the influence of external bias. To this end we will employ the Ginzburg-Landau 
(GL) equation with the DeGennes surface condition where only the latter condition depends on the external bias in agreement with
the Anderson theorem.
%, the GL equation is not affected by the 
%applied electric field. The gate voltage changes only the surface
%condition via dependence of the surface density of states on the
%induced surface charge.

\begin{figure}
\centerline{\psfig{file=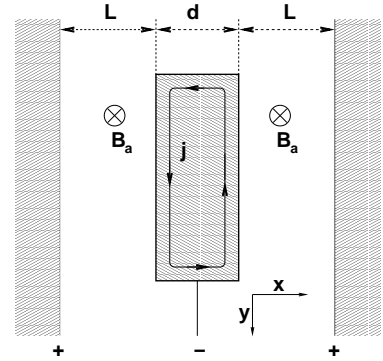,width=5cm}}
\caption{The slab superconductor of thickness $d$ placed in two parallel capacitor plates at distance $L$. The superconductor extends infinitely in z direction. The magnetic field is parallel to the superconductor surface. 
\label{exsetup}}
\end{figure}

The experimental setup is 
shown in figure~\ref{exsetup}.
The first electrode of a capacitor is the superconducting slab 
of thickness $d$. It is sandwiched between two plates of an ideal
metal at distances $L$, which form the second electrodes. In real devices it will 
be necessary to consider that the applied electric field affects 
also the end corners. Here we neglect the influence of the 
end corners assuming an 
infinite plane capacitor. Our aim is to evaluate the effective capacitance, the surface critical field and the surface energy in dependence on the applied voltage and the magnetic field.
 
The total energy of the capacitor with area $S$ is given by $\frac 1 2 \epsilon_0 E^2 L S$ and an additional contribution coming from surface charges $S \sigma$. Then the inverse capacitance $C$ of the slab is given by
\be 
{S\over C}&=&{L\over \epsilon_0}+{1\over \epsilon_0^2}{\partial^2 \sigma\over \partial E^2}.
\label{capac}
\ee
The external magnetic field associated with the $z$ axis and parallel to the superconducting surface is screened inside the superconductor by the diamagnetic current ${\bf j}$. Thus a magnetic field profile is established perpendicular to the surface induced by the diamagnetic current. This induced diamagnetic current
%responsible for the Meissner effect is the dominant 
contributes to the inverse capacitance such that we have besides a genuine surface contribution $C_{\rm surf}$ also an induced part  $C_{\rm ind}$,
\be 
{1\over \epsilon_0^2}{\partial^2 \sigma\over \partial E^2}&=&{S\over C_{\rm surf}}+{S\over C_{\rm ind}}.
\ee
It will turn out that the induced capacitance is linearly dependent on the sample width $d$.

Since the electric field penetrates the superconductor only near the surface it is of special interest to understand the surface superconductivity in the presence of an external electric field. The change of the upper critical field and the surface nucleation field has been calculated already for strong coupling.\cite{EA67}  The surface paraconductivity and the change of the critical parameters due to an external field has been investigated, too.\cite{Sh85,Sh93} A shift of the critical temperature has been obtained \cite{ShK93} due to a modified GL boundary condition and a variational solution of the effective Schr\"odinger equation. In other words, the critical field $B_{c1}$ is changed due to the change of the GL energy by the electric field.\cite{SKB94} Here we will investigate the surface energy problem of domain walls similarly and will employ the modified boundary condition to solve the GL equation variationally. We will obtain that the bulk critical field $B_{\rm c2}$ remains unchanged due to the applied electric field while the surface critical field $B_{\rm c3}$ changes with the electric field.

The paper is organized as follows. First we repeat the solution of the magnetic-field dependent GL equation under external bias by electric fields and calculate the surface critical field $B_{\rm c3}$ and its dependence on the external bias. With the help of the GL wave function the surface energy is calculated in chapter III and the effective capacitance in chapter IV. Special attention is paid to the induced magnetic field effect in chapters III and IV. The selfconsistent treatment of the induced magnetic field is presented in appendix~\ref{bint} which completes the proposed picture. We present fitting formulas for the magnetic and electric field dependence aimed for experimental verifications. In chapter V we summarize and discuss possible experimental realizations. 

\section{GL wave function with external magnetic and electric fields}

An effective description of superconducting properties near the critical temperature is provided 
by the GL equation for the wave function $\Psi$, 
\be
{1\over 2 m}\left(-i\hbar\nabla-e{\bf A}\right)^2\Psi+
\alpha\Psi+\beta|\Psi|^2\Psi=0,
\label{e}
\ee
 which describes the ratio of the superconducting density to the total density $n$ by $|\Psi|^2=n_s/2 n$. Here the mass is twice the electron mass, $m=2 m_e$, and the charge is $e=2 e_e$, the one of the Cooper pairs. If needed, the
effective potential can be extended to lower temperatures than the critical 
one.\cite{B54,LKMBY07} 

The GL equation is supplemented by the 
DeGennes surface conditions\cite{dG89}
\be
\left.{\nabla\Psi\over\Psi}\right |_{x=0}={1\over b},
~~~~~~~~~~~~~
\left.{\nabla\Psi\over\Psi}\right |_{x=d}=-{1\over b},
\label{b1}
\ee
where the extrapolation length $b$ is sensitive to the treatment
of the surfaces. 
%The vector potential is given by the Ampere law
%\be
%\nabla^2{\bf A}=-\mu_0{e^*\over m^*}\bar\Psi\left(-i\hbar\nabla-
%e^*{\bf A}\right)\Psi.
%\label{Amp1}
%\ee
This inverse extrapolation length $1/b$ depends on the density of
states at the surface, therefore it is a function of the applied
electric field $E$. In linear approximation, see appendix~\ref{boundc}, 
it reads
\be
{1\over b}={1\over b_0}+{E\over \varphi_{\rm fe}},
\label{b1a}
\ee
with the characteristic potential\cite{LMKY06} 
\be
{1\over \varphi_{\rm fe}}=
{4e\over mc^2}~\kappa^2~\eta~
{\partial\ln T_{\rm c}\over\partial\ln n}
\label{Pavel}
\ee
being of the order of few MeVs for conventional superconductors.
The dependence of $1/\varphi_{\rm fe}$ on the GL parameter $\kappa$ 
and the density derivative of the critical temperature suggests 
that the field effect is much larger for high-$T_{\rm c}$ superconductors.

The reduction factor $\eta$ is the ratio of the gap extrapolated 
to the surface and the value at the surface.\cite{LMKY06} Its
value is of the order of unity and it is not essential for our
discussion.

\subsection{Nucleation of the surface superconductivity}

At the surface critical field the superconductivity nucleates at
the surface. Near the surface the effective wave function $\Psi$ is small and we can work with the linearized GL equation, omitting in (\ref{e}) the cubic term,
\be
{1\over 2 m} (i\hbar \nabla-e{\bf A})^2  \Psi+\tilde \alpha  \Psi=0
\label{E1}
\ee
with the boundary condition (\ref{b1}).
%, $\nabla \tilde\Psi/\tilde\Psi|_0=1/b$. 

We consider the geometry of a planar superconductor at $d>x>0$ as in figure~\ref{exsetup} and assume a homogeneous
applied magnetic field ${\bf B}_{\rm a}=(0,0,B_{\rm a})$. Since
the system in Fig.~\ref{exsetup} has translation invariance along 
the $y$ direction, we use the Landau gauge of the form 
\be
{\bf A}=(0,B_{\rm a} x,0).
\label{a}
\ee 

The nucleation is possible if the parameter $-\tilde \alpha$ of (\ref{E1}) 
becomes equal 
to an eigenvalue $\varepsilon$ of the kinetic energy given by
${1\over 2 m}\left(-i\hbar\nabla-e{\bf A}\right)^2\psi=
\varepsilon\psi$. Since $\alpha$ changes with the temperature,
$\alpha=\alpha'(T-T_{\rm c})$, the eigenvalue $\varepsilon$ of the 
kinetic energy determines the nucleation temperature $T^*$ as 
$T^*-T_{\rm c}=-\varepsilon/\alpha'$.\cite{LMKY06} To avoid dual notation for
the same quantity, we will treat the equation (\ref{E1}) as an eigenvalue problem for 
$\tilde \alpha$.
% and evaluate the nucleation temperature as 
%\be
%T^*=T_{\rm c}+{\alpha\over\alpha'}.
%\label{Tnuc}
%\ee
Since $\tilde \alpha$ is negative, the nucleation temperature $T^*$ is 
always below the critical temperature $T_{\rm c}$ in the absence 
of the magnetic field.

Assuming the translation invariance along the $y$ and $z$ axes we 
can write the wave function as
\be
\Psi(x,y,z)=\psi(x)~{\rm e}^{iky}{\rm e}^{iqz}.
\label{wavef}
\ee
Using (\ref{wavef}) in the GL equation (\ref{E1}) we get a 
one-dimensional equation
\begin{align}
{\hbar^2\over 2 m}\left(-\left({\partial\over\partial x}\right)^2+
\left(k-{eB_{\rm a}\over\hbar}x\right)^2+
q^2\right)\psi+
\alpha\psi=0.
\label{E1b}
\end{align}

Any non-zero value of $q$ results in the kinetic energy $q^2\hbar^2/2m$ 
which lowers the value of $\alpha$ making the nucleation temperature 
lower. The nucleation happens on the first possible occasion, i.e., at
the highest allowed temperature. We thus take $q=0$.

The value of $k$ determines 
the minimum of the parabolic potential and the eigenvalue $\alpha$
depends on the relative position of this minimum with respect to the
surface of the slab. We have to find the wave vector $k$ from the requirement 
of the highest nucleation temperature. 

%Then the wave function assumes the form 
%\be
%\tilde \Psi(x,y)=\psi(x) \exp{(i k y)}
%\label{wavef}
%\ee
%where the momentum $k$ will be determined by the requirement that the ground state should correspond to the lowest eigenvalue.

\subsection{Thick slab limit}
First, we assume that the superconducting slab is so thick that its
surface superconductivity forms on both surfaces independently.
In this case we can view the sample as infinite and take
the convergent solution into the bulk. We treat only the surface
at $x=0$. The surface at $x=d$ is analogous. We should note that 
the profile for a general thickness has been solved by a calculation 
based on the Eilenberger equation for finite temperatures but without 
bias voltage.\cite{SP91} Here we restrict to a simpler approach but have 
included the bias voltage.

It is advantageous to express the $x$-coordinate with the help of 
the dimensionless coordinate $\tau$
\be
x=\tau l +2 l^2 k,
\label{coordinates}
\ee
such that the wave function reads
\be
\psi(x)={\cal C} D_{\tilde\nu}\left ({x\over l}+\tau_0\right )
\label{wave}
\ee 
with the momentum $\tau_0=-2 k l$ and the magnetic length 
\be
l^2={\hbar \over 2 eB_a}.
\label{l}
\ee 
The parabolic cylinder function $D_{\nu}(\tau)$  solves the differential equation (\ref{E1}), i.e. \cite{MOS66}
\be
{d^2 D_\nu(\tau)\over d\tau^2}&=&\left ({\tau^2\over 4}-\nu-\frac 1 2\right )D_\nu(\tau) \label{dfg}\\
\left . {D_\nu'(\tau_0)\over D_\nu(\tau_0)}\right |_{\tau_0=-k\sqrt{2 \hbar \over e B_a}}&=&{l \over b}, \label{bound}
\ee
with 
\be
\nu=-\frac 1 2 -{\alpha m\over e \hbar B_a}
\label{nu}
\ee
and the GL coherence length  $\xi^2=-\hbar^2/2 m \alpha$.

\begin{figure}[h]
\centerline{\psfig{file=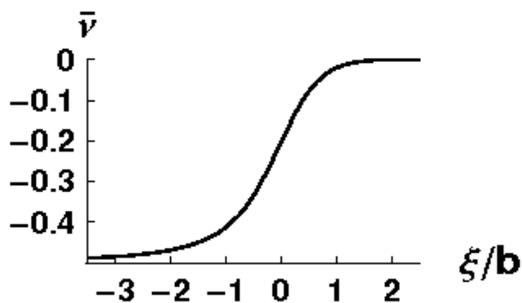,width=7cm}}
\caption{The minimal Eigenvalue of the GL equation given by (\ref{bsol}) as a function of external bias.}
\label{nu_b}
\end{figure}

The boundary condition (\ref{bound}) leads to a function $\nu(\tau_0)$. The maximal nucleation temperature is given by the maximal $\tilde \alpha={\rm max}[\alpha]$ which is characterized by the minimum $\tilde \nu={\rm min}[\nu(\tau_0)]$ due to (\ref{nu}). 
Besides the obvious numerical search we can give directly a nonlinear equation for this desired minimum  $\tilde\nu'(\tau_0)=0$. For this purpose we differentiate (\ref{bound}) with respect to $\tau_0$, using the relations for the parabolic cylinder functions, $D'_\nu=\tau D_\nu/2-D_{\nu+1}$ and $D_{\nu+1}=\tau D_\nu-\nu D_{\nu-1}$, to arrive at
\be
&&\left . {D_{\tilde\nu+1}(\tau_0)\over D_{\tilde\nu}(\tau_0)} \right |_{\tau_0
=-2 \sqrt{(\tilde\nu+\frac 1 2 )(1+{\xi^2\over b^2})}}\nonumber\\&&\qquad=-\sqrt{{\left (\tilde\nu+\frac 1 2 \right )\left (1+{\xi^2\over b^2}\right )}}-{\xi \over b}\sqrt{\tilde\nu+\frac 1 2}.
\label{bsol}
\ee 
With the solution $\tilde \nu[\tau_0]$  of (\ref{bsol}) the momentum and current is determined due to $\tau_0=-2 k l$. In figure~\ref{nu_b} the solution of (\ref{bsol}) is plotted. There is an asymmetry to be noticed with respect to positive, $b>0$, and negative, $b<0$, external bias. This will lead to very asymmetric curves in the surface energy later.

\subsection{Surface critical field}
The lowest eigenvalue $\tilde \nu={\rm min}[\nu(\tau_0)]$ of (\ref{dfg}) corresponds to the highest attainable critical magnetic field
\be
B_{\rm c3}={\rm max} (B_a)={-m \alpha\over \hbar e (\tilde \nu+\frac 1 2)}\equiv {B_{\rm c2} \over 2 \tilde \nu +1}
\label{bc3}
\ee
where $B_{\rm c2}$ is the upper critical field.  The modified boundary condition does not influence $B_{\rm c2}$ but makes it possible that $-1/2<\tilde \nu \le 0$ and a higher critical magnetic field $B_{\rm c3}>B_{\rm c2}$ appears such that the superconductivity near the surface is enhanced in dependence on the external electric field.

\begin{figure}[h]
\psfig{file=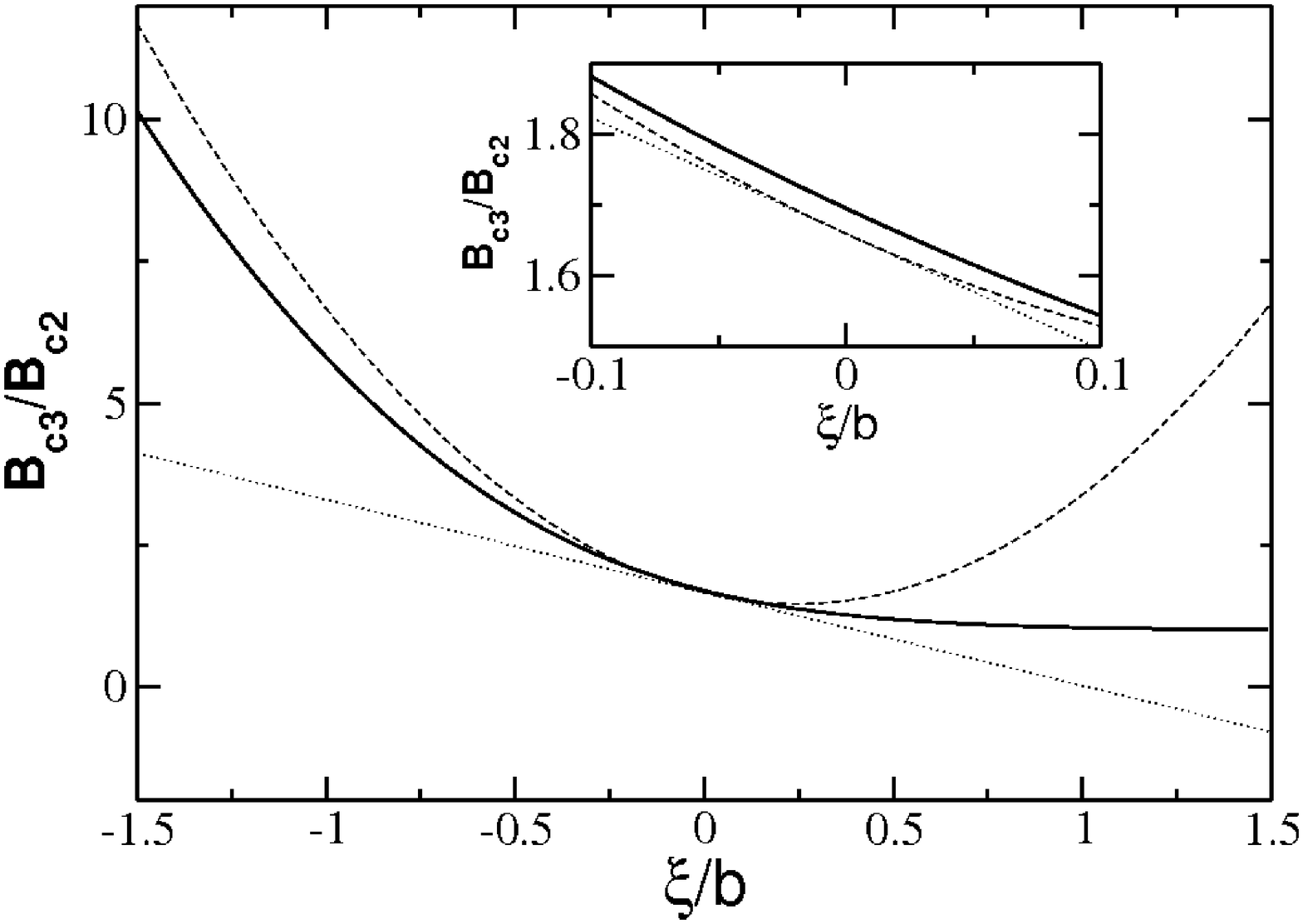,width=9cm}
\caption{The surface critical field $B_{\rm c3}$ versus the external bias (\ref{b1}) from linearized GL equation. The solution (\ref{bc3}) (solid line) is compared with the expansion up to first (dotted line) and second (dashed line) order in $1/b$ according to (\ref{16}).
% For comparison the line of vanishing surface energy is given as well (thick dotted line).
}
\label{bc3f}
\end{figure}

In figure \ref{bc3f} we present the result for the surface critical field (\ref{bc3}) versus the external bias (\ref{b1}). We see that the external bias can enhance or decrease the surface critical value depending on the field direction. Without external bias the known GL solution $B_{\rm c3}/B_{\rm c2}=1.69461$ is reproduced.\cite{SJdG63} The strong coupling limit is somewhat larger resulting in the value $1.8$. \cite{EA67} We see that the electric field can generate easily a value larger than $1.695$. The experimental values compared with the GL theory and the theory of Hu and Korenman \cite{HK69} are discussed in \cite{Ki75} which shows that the GL values are too small. 

To provide analytical expressions let us remind a simpler variational solution of the problem which was contributed by Kittel for absent external bias.\cite{dG89,Tinkham}. We can extend this approach to external voltage bias and choose for the wave function the ad hoc ansatz $\psi\approx {\cal C} \exp{(-a\, x^2+x/b)}$ which obeys the DeGennes boundary condition $\psi'/\psi|_0=1/b$ automatically. The constants $a$ and $\tau_0=-2 k l$ have to be determined by the minimal eigenvalue of (\ref{E1}). The normalized minimal mean eigenvalue $<\tilde \alpha>$ can be obtained by minimizing the functional 
\be
<\tilde \alpha>={\int\limits_0^\infty d x \left [{l_{\rm min}^{-4}} (x+\tau_0 l)^2 \phi^2
%+{2 \over b} \phi' \phi
-\phi'' \phi \right ]\over \int\limits_0^\infty d x \phi^2}
\label{fun}
\ee
with respect to $\tau_0$ and $a$. The resulting $l_{\rm min}^{-2}=2 e B_{\rm c3}/\hbar$ yields
\be
{B_{\rm c3}\over B_{\rm c2}} &\equiv& {\xi^2 \over l_{\rm min}^2}=
\sqrt{\frac{\pi }{-2+\pi
   }}-\frac{2
   }{(-2+\pi )^{3/2}}{\xi\over b}\nonumber\\
&
+&\frac{(21+2 \pi  (2+(-4+\pi ) \pi )) }{2
   (-2+\pi )^{5/2} \sqrt{\pi }}{\xi^2\over b^2}+{\cal O}({1\over b^{3}}).
\label{16}
\ee
In the case of vanishing electric fields corresponding to the boundary condition $1/b\approx 0$ we recover the known results
\cite{Tinkham}, $a={1/ 2 \xi^2}$, $x_0={1/ \sqrt{2\pi a}}$ and ${B_{\rm c3}/ B_{\rm c2}}=\sqrt{\pi/( \pi-2)}\approx 1.66$. The comparison of the expansion (\ref{16}) with the solution (\ref{bc3}) can be seen in figure~\ref{bc3f}.

\subsection{Variational wave function}
Let us now return to the full solution of (\ref{E1b}). The wave function (\ref{wave}) specified by the value $\nu=\tilde \nu$ describes the situation for the maximal nucleation temperature corresponding to the maximal magnetic field $B_a\approx B_{\rm c3}$. 
 
\begin{figure}[h]
\psfig{file=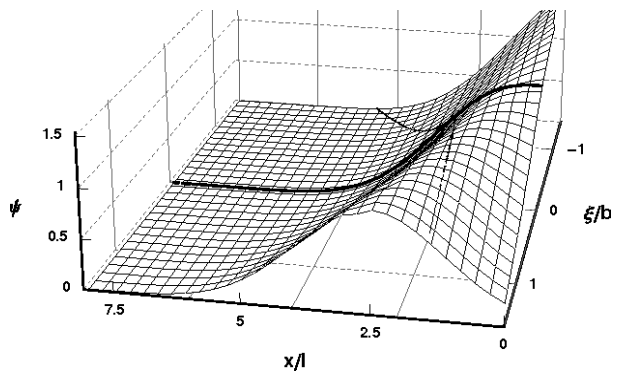,width=8cm}
\caption{The GL wave function of the condensate versus the external bias (\ref{b1}). The normal GL wave function without external bias is marked as thick line. The transition curve of $\xi/l$ at $B_a=B_{\rm c3}$ according to (\ref{21}) is plotted as dotted line.}
\label{psi}
\end{figure}

In figure~\ref{psi} we see how the external bias, $1/b$, changes the wave function in dependence on the distance of the surface. The magnetic fields enters merely as a scaling of the spatial coordinate.  For positive electric fields the superconducting density is diminished on the surface while for oppositely directed electric fields the surface superconductivity is enhanced.
 
Since the wave function is strictly valid only near the transition line $B_a\approx B_{\rm c3}$ we have plotted in figure~\ref{psi} also the transition line 
\be
\left .{\xi\over l}\right |_{B_a=B_{\rm c3}}={1\over \sqrt{\tilde \nu+\frac 1 2}}
\label{21}
\ee 
for $x=\xi$ as orientation.

We will use the wave function (\ref{wave}) as an ansatz for the variational calculation of the surface energy in such a way that the amplitude ${\cal C}$ serves as a variational parameter. This is motivated by the fact that near the surface the shape of the wave function is only slightly changed compared to the surface values but the amplitude decreases exponentially away from the surface.

\section{Surface energy}
With the help of the GL wave function we can now calculate the surface energy. 
This surface energy is the integral over the energy difference between the actual Gibbs free energy and the Gibbs free energy deep in the superconductor. Since the latter equals the one deep in the  normal region when the field energy is subtracted, $G(x\to -\infty)=G_{n0}-{B_a^2/ 2\mu_0} = G_{s0}$, we can write the surface energy as 
%\footnote{I neglect in the following the $E(z)$ profile. This should be included, but I don't know presently. It will lead to modifications to the final results. Further to note that we optimized the variational wave function to the interval $(0,\infty)$ such that the integrations
%should be limited to that range too.}
\be
&&\!\!\!\!\!\sigma\!=\!\int\limits_{0}^\infty \!d x \left [G(x)-G_{n0}+{B_a^2\over 2 \mu_0}\right ]
=\int\limits_{0}^\infty \!d x \left [G(x)-G_{s0}\right ]
%\nonumber\\\!&=&\!
%\int\limits_{-\infty}^\infty \!d z
%\left [F(z)\!-\!B(z) H_c\!-\!D(z) E\!-\!F_{n0}\!+\!{B_c^2\over 2\mu_0}\!+\!{E^2\over 2 \epsilon_0}
%\right ]
\nonumber\\
&&\!\!\!\!\!=\!\!
\int\limits_{0}^\infty \!\!d x \!\left 
[\alpha |\Psi|^2\!+\!{\beta \over 2} |\Psi|^4\!+\!{[B_a\!-\!B(x)]^2\over 2 \mu_0}
%\right . 
%\nonumber\\&&
%\left .
\!+\!{|(i\hbar {\bf \nabla}\!+\!e {\bf A})\Psi|^2 \over 2 m}
\right ]
\nonumber\\&&
+{\hbar^2\over 2 m} \Psi(0)\Psi'(0).
\label{sur}
\ee
The last counterterm is necessary in order to provide a consistent variational problem with the modified GL boundary condition (\ref{b1}), for details see appendix~\ref{paradox}.

The surface energy appears only if terms $\sim |\Psi|^4$ are taken into account. Since the shape of the GL wave function changes much less in terms of the applied magnetic field than the amplitude, we can now use our solution of the linearized equation, $\Psi={\cal C} D_{\tilde\nu}$, to calculate the surface energy. For this purpose we determine the constant $N$,
\be
{\cal C}^2=N{\alpha\over \beta}
\label{calc}
\ee 
such that (\ref{sur}) takes a minimum. 

\begin{figure}
\psfig{file=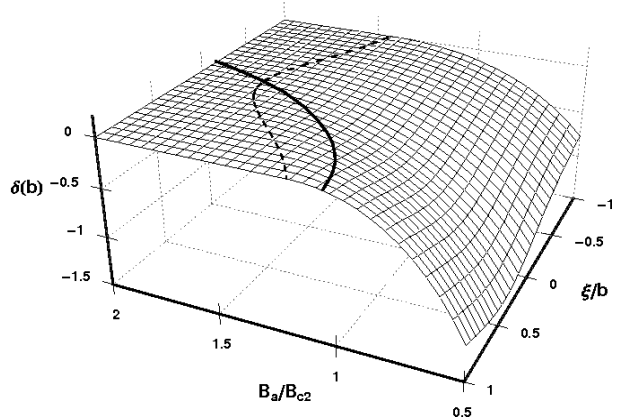,width=8cm}
\caption{The surface energy in terms of the wall parameter (\protect\ref{wallp}) versus the magnetic field and the external bias $1/b$. The solid line denotes the transition curve $B_a=B_{\rm c3}$ according to (\ref{21}) 
% and the dotted line the zero of 
where the surface energy vanishes.}
\label{surf}
\end{figure}

\subsection{Limit of thick samples}

Assuming the limit of thick sample, $d\gg l$, we neglect in the first step the 
space profile of the magnetic field near the surface. This profile or induced field effect due to diamagnetic currents will be discussed separately in the next paragraph. In the same
spirit, the upper integration limit is taken as infinite here at this moment. 
The nonlinear Gibbs free energy (\ref{sur}) reads then
\be
G&=&G_0-l {B_c^2\over 2 \mu_0} \left ( 2 N {\cal A}-N^2 {\cal B} \right )
%\!+\!{\cal C}^2 l\! \int\limits_{\tau_0}^\infty \!\! d\tau \left [ \left (\alpha +{\tau^2 \hbar^2 \over 8 m l^2} \right ) D_{\tilde \nu}^2(\tau)+ {\hbar^2 \over 2 m} D_{\tilde\nu}'^2(\tau)\right ]
%\nonumber\\&+&
%{{\cal C}^4 \beta \over 2} \int \limits_{\tau_0}^\infty D_{\tilde \nu}^4(\tau).
\label{f}
\ee
Here we have introduced
\be
{\cal A}&=&\int\limits_{\tau_0}^\infty d \tau \left [ \left ( {\tau^2 \xi^2 \over 4 l^2}-1\right ) D_{\tilde\nu}^2(\tau)+{\xi^2\over l^2}  [D'_{\tilde\nu}(\tau)]^2\right ]\nonumber\\&&
+{\xi^2\over l^2} D_{\tilde \nu}(0)D_{\tilde \nu}'(0)
\nonumber\\
&=&-\left (1-{\tilde \alpha \over \alpha}\right ) \int\limits_{\tau_0}^\infty d\tau D_{\tilde \tau}^2(\tau)
\nonumber\\
{\cal B}&=&\int\limits_{\tau_0}^\infty d \tau  D_{\tilde\nu}^4(\tau)
\label{ab}
\ee
and used $\alpha^2/\beta=B_c^2/\mu_0$. The minimum of (\ref{f}), 
\be
N={\cal A}/{\cal B},
\label{cc}
\ee
%\be
%{\cal C}^2={\alpha{\cal A}\over \beta{\cal B}}
%\label{cc}
%\ee 
leads to a surface energy $\sigma$ in terms of the condensation energy expressed in the critical field
\be
\sigma&=&{B_c^2\over 2 \mu_0} \,\,\delta(b)
\label{ss}
\ee
with the wall parameter 
\be
\delta(b)= -l {{\cal A}^2\over {\cal B}}.
\label{wallp}
\ee 
In figure~\ref{surf} the wall parameter is plotted versus the external bias 
and the magnetic length. We see that with increasing magnetic fields, i.e. decreasing magnetic length, the negative surface energy increases dependent on the external bias. Therefore the surface energy can be changed by the applied magnetic field as well as the external bias. 
The line of minimal eigenvalues of the GL equation (\ref{21}) called transition curve is shown as well where $B_a=B_{\rm c3}$ and the surface energy vanishs. 

It is instructive to derive the wall parameter for the case without external bias,
\be
\delta(\infty)&=&-l \left [\!{\xi^2\over l^2} \!\left (\tilde \nu\!+\!\frac 1 2 \right )-1 \! \right ]^2 \!\!{\cal D}
\nonumber\\
&=&-\lambda  \sqrt{{B_c\over B_a}}{\left [(2 \tilde \nu\!+\!1)\sqrt{{B_a\over B_c}}\!-\!\sqrt{2{B_c\over B_a}} \kappa \right ]^2 \over (\sqrt{2} \kappa)^{5/2}}{\cal D}
\label{del00}
\ee
with 
\be
{\cal D}={\left (\int\limits_{\tau_0}^\infty d\tau D_{\tilde \nu}^2(\tau) \right )^2\Big/ \int\limits_{\tau_0}^\infty d\tau D_{\tilde \nu}^4(\tau)}.
\ee
The upper critical field is related to the GL parameter $B_{\rm c2}=\sqrt{2} \kappa B_{\rm c}$ and $l^2/\xi^2=B_{\rm c2}/2B_a$. With the help of (\ref{bc3}) it is also easy to check that the surface energy (\ref{del00}) is exactly zero for $B_a=B_{\rm c3}$ in the case of absent external bias.

Assuming magnetic fields $B_a\approx B_c$ in order to adapt to the situation of the superconductor-normal surface wall parameter we obtain
\be
\lim\limits_{B_a\to B_c} \delta(\infty)
=-\lambda {2.032\over \kappa^{5/2}} ( 0.41727-\kappa)^2.
\label{del26a}
\ee
An approximate treatment of the wall parameter of the superconductor-normal region is presented in appendix~\ref{domain}.  Neglecting induced diamagnetic currents the wall parameter $\delta$ for both type-I and type-II superconductors \cite{Tinkham,FW71,dG89} is approximated by (\ref{del0a})
\be
\delta_{\rm s-n}\approx -\lambda \left (\frac 3 2-{4 \sqrt{2} \over 3 \kappa} \right )
\label{deltadG}
\ee
in terms of the London penetration depth $\lambda$ of the magnetic field and the GL parameter $\kappa=\lambda/\xi$ with the coherence length $\xi$. 

\begin{figure}
\psfig{file=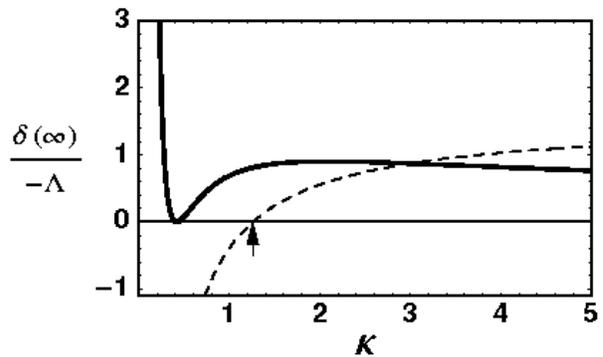,width=8cm}
\caption{The surface wall parameter (\protect\ref{del00}) for superconductor-vacuum boundary and zero external bias (\ref{del26a})  versus the GL parameter (solid line) compared to the superconductor-normal boundary expression (\ref{deltadG}) of the literature (dotted line).}
\label{del0}
\end{figure}

The result (\ref{del26a}) for the superconductor-vacuum transition can be compared with this superconductor-normal boundary (\ref{deltadG}).
In figure~\ref{del0} we plot (\ref{del26a}) and (\ref{deltadG}) and one sees the different places where the surface energy is vanishing. This vanishing of surface energy is connected with the transition from type-I to type-II superconductivity. In the latter case the surface energy is negative indicating an unstable surface forming a vortex structure. While the superconductor-normal result (\ref{deltadG}) leads to 
\be
\kappa_0|_{\rm s-n}=8\sqrt{2}/9=1.257
\ee 
the superconductor-vacuum result (\ref{del26a}) suggests a smaller value 
\be
\kappa_0|_{\rm s-v}={2 \tilde \nu+1\over \sqrt{2}}=0.41727.
\ee 
Without external magnetic field $\tilde\nu=0$ the superconductor-vacuum result (\ref{del26a}) coincides with the transition point between type-I and type-II superconductivity. 
In other words the superconductor-vacuum boundary leads to smaller values of the transition between type-I and type II than the superconductor-normal boundary with respect to the stability of the surfaces. We find that the applied magnetic field decreases the transition GL parameter. The type-II superconductivity extends towards values below $\kappa=1/\sqrt{2}=0.7071$.

\subsection{Finite width of samples}

When the applied magnetic field $B_a$ exceeds the upper critical field $B_{\rm c2}$ no bulk superconductivity is possible anymore but in a small surface region the surface superconductivity occurs up to the surface critical field $B_{\rm c3}$. This critical field is dependent on the thickness of the sample \cite{SP91}. We have discussed the thick sample limit with respect to London's penetration depth, $d\gg \lambda$, so far. Now we are going to consider the finite sample limit.

Investigating a layer of finite thickness, $d<\infty$ the Gibbs free energy (\ref{f}) with (\ref{ab}) possesses an upper integration limit $d/l+\tau_0$ instead of $\infty$. The superconducting surfaces are then also characterized by the appearance of diamagnetic currents. These diamagnetic currents induce magnetic fields, see appendix~\ref{bint}. These fields contribute to the surface energy. We solve again the variation problem of the free Gibbs energy (\ref{f}) in order to obtain the optimum ${\cal C}$. The expression for the surface energy (\ref{sur}) remains the same but the magnetic profile is now spatial dependent,
\be
B(x)=B_a+\mu_0 M(x).
\ee 
The external magnetic field cancels in the difference of the Gibbs energies and is present only in the vector potential which takes now the form (\ref{ay}),
\be
A_y(x)=B_a x+\mu_0 \int\limits_0^x dx' M(x'),
\ee
instead of (\ref{a}).  As a result, (\ref{f}) assumes
\be
G&=&G(M\!=\!0)\!-l {B_c^2\over 2 \mu_0} \left \{
\!{N\over \sqrt{2}\kappa}\! \int\limits_{\tau_0}^{d/l+\tau_0} \!d\tau D_{\tilde \nu}(\tau)^2 
\right .\nonumber\\&&
\left .
\times\left [
{B_c\over B_a} \left (\int\limits_{\tau_0}^\tau \! d\tau' {\mu_0 M(\tau')\over B_c}\right )^2    
+
2 \tau \int\limits_{\tau_0}^\tau \! d\tau' {\mu_0 M(\tau')\over B_c}
\right ]
\right .\nonumber\\&&
\left .
-\int\limits_{\tau_0}^{d/l+\tau_0} \!d\tau \left ({\mu_0 M(\tau)\over B_c}\right )^2
\right \}
\nonumber\\
&=&\!-{l B_c^2\over 2 \mu_0} \biggl [ 2 N {\cal A} \!-\!N^2 ({\cal B}\!+\!{\cal B}')\!+\!{\cal D'} N^3  \biggr ]\!+\!{\cal O}(N^4).
\label{gnn}
\ee
Here we have used (\ref{Mn}) in the last line. Besides (\ref{ab}) we obtain now additional integrals, ${\cal B}'$ given by (\ref{bs}) and ${\cal D}'$ by (\ref{ds}) calculated in appendix~\ref{integc}. 

The Gibbs free energy (\ref{gnn}) contains now terms $\sim {\cal C}^6$ and the minimization yields a quadratic equation for ${\cal C}^2$. Instead of (\ref{cc}) we have
\be
{N}={{\cal A}\over {\cal B}+{\cal B}'}{1\over {\cal Y}} \left (1-\sqrt{1-2 {\cal Y}}\right )
\label{nn}
\ee
with 
\be
{\cal Y}=3 {{\cal A} {\cal D}'\over ({\cal B}+{\cal B}')^2}.
\label{yy}
\ee
The resulting surface energy takes the form of (\ref{ss}) with the wall parameter
\be
\delta_{\rm ind}(b)=-l {{\cal A}^2\over {\cal B}+{\cal B}'} g(\cal Y)
\label{walld}
\ee
and
\be
g(x)={2\over 3 x^2}\left [3 x-1+(1-2 x)^{3/2}\right ]\approx 1+{x\over 3}+... .
\ee

\begin{figure}
\psfig{file=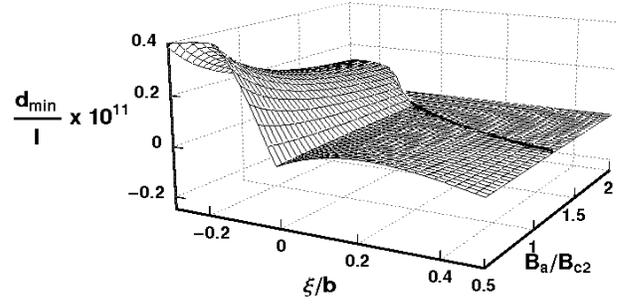,width=8cm}
\caption{The minimal thickness of the sample considered within the present approach versus the applied magnetic field and the external bias $1/b$ in units of the coherence length $\xi$. The transition line according to (\ref{21}) is given as well.}
\label{dmin}
\end{figure}

From (\ref{nn}) we see that the variational solution is only meaningful if ${\cal Y}<1/2$. This specifies the lower limit on the thickness $d_{\rm min}$ of our sample we can consider within this approximation. From (\ref{yy}) and with the help of (\ref{bs}) and (\ref{ds}) we obtain
\be
d_{\rm min}> {12 {\cal A} {\cal J} \over {\cal F}^3} {l^3\over \xi^2} +{\cal O} (d^{-1})
\ee
where ${\cal F}$ given by (\ref{fvw}) and ${\cal J}$ by (\ref{jj}). 
This lower limit is plotted in figure~\ref{dmin} in terms of the magnetic length. One sees that we have practically a visible lower limit only for strong negative values of the external bias.

\subsubsection{Limit towards large sample thickness}

It is now interesting to discuss the limit of thick superconductor probes. As shown in appendix~\ref{integc}, both expressions ${\cal B}'$ and ${\cal C}'$ are linearly dependent on the thickness $d$. We obtain for the wall parameter (\ref{walld})
\be
&&\delta_{\rm ind}(b)=-{2 \kappa^2 l^2 {\cal A}^2\over {\cal F}^2 d}+{\cal O}(d^{-2})
\label{dll}
\ee
where ${\cal F}$ is given by (\ref{fvw}). In the case of absent external bias one has
\be
&&\delta_{\rm ind}(\infty)=-{2 \kappa^2 l^2\over d \langle x \rangle ^2} \left [{\xi^2\over l^2} \left (\tilde \nu+\frac 1 2 \right )-1 \right ]^2
\nonumber\\
&&=  -{\lambda \xi\over d \langle x \rangle ^2\sqrt{2} } \left [(2 \tilde \nu +1) \sqrt{{B_a\over B_c}}- \kappa \sqrt{2{B_c\over B_a}} \right ]^2 
\label{walll}
\ee  
with $\xi^2/l^2=2 B_a/B_{\rm c2}$, $B_{\rm c2}=\sqrt{2}\kappa B_c$ and the mean distance
\be
\langle x \rangle={\int\limits_{\tau_0}^{d/l+\tau_0} \!\!\!\!d\tau \tau D^2_{\tilde \nu}(\tau)\Big / \int\limits_{\tau_0}^{d/l+\tau_0} \!\!\!\!d\tau D^2_{\tilde \nu}(\tau)}.
\ee
If $B_a\approx B_c$ we have
\be
\lim\limits_{B_a\to B_c} \delta_{\rm ind}(\infty)=
  -744326 \,\, {\lambda \xi\over d }\,\,(0.41727-\kappa)^2.
\ee
We see that the surface energy vanishes at the same GL parameter as without induced fields (\ref{del26a}).

\begin{figure}
\psfig{file=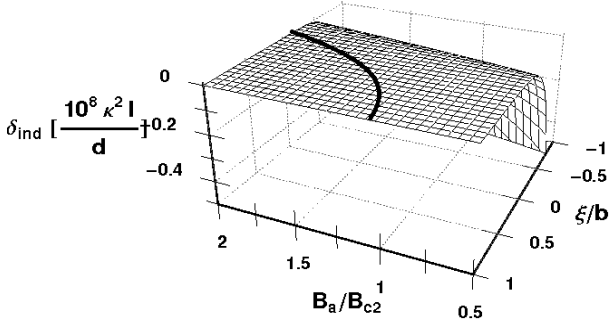,width=8cm}
\caption{The induced part of the surface energy (\ref{dll}) in terms of the wall parameter (\protect\ref{wallp}) versus the magnetic field and the external bias $1/b$. The solid line denotes the transition curve $B_a=B_{\rm c3}$ 
%and the dotted line the zero of the surface energy
.}
\label{surfind}
\end{figure}

\section{Effective capacitance}

Now we return to the experimental setup of figure~\ref{exsetup} and determine the expected contributions to the effective capacitance (\ref{capac}). Besides the ideal capacitance, $C_0=\epsilon_0 S/L$ we obtain contributions due to the external bias, 
\be
&&{S\over C_{\rm ex}}={1\over \epsilon_0^2}{\partial^2 \sigma\over \partial E^2}
= {\xi^2\over \epsilon_0^2 \varphi_{\rm el}^2}{\partial^2 \sigma \over \partial \left ({\xi/ b}\right )^2} 
\nonumber\\
&&={L_0\over \epsilon_0} {l\over \xi} {\partial^2  \over \partial \left ({\xi/ b}\right )^2}\left (\!{\delta\over l}\!\right )4 (1\!-\!t)^2
=4 (1\!-\!t)^2\left. {S\over C_{\rm ex}}\right |_0 .
\label{dl}
\ee
where  the temperature dependence $t=T/T_c$ arises from $B_c^2/2\mu_0=(\epsilon_c/n)(1-t^2)^2$ and $\kappa^2(T)=2 \kappa^2\,/(1+t^2)$ in (\ref{Pavel}) and we have scaled with respect to the temperature-independent coherence length $\xi=\xi(T) \sqrt{1-t^2}$ and $\varphi_{\rm el}$ of (\ref{Pavel}).
We abbreviate 
\begin{align}
{L_0\over \epsilon_0}={\xi^3\over \epsilon_0^2 \varphi_{\rm el}^2} {B_c^2\over 2 \mu_0}=
{\epsilon_c \over n e \kappa_0^3} \left ({m c^2\over e}\right )^{3/2}\!\!\!\!\!\!\!{1\over\sqrt{n e \epsilon_0} } {1\over \varphi_{\rm el}^2|_{t=0}}.
\label{L0}
\end{align}

Using the BCS expression for the GL parameter $\beta= 24 \hbar^2/( 7\zeta[3]  n m 1.76^2 \xi_{\rm BCS}^2)$ we can rewrite (\ref{L0}) also into another form
\be
{L_0\over \epsilon_0}= \alpha_0^4  {7 \pi \zeta[3] \over 6} a_B^3 n \left (\! 1.76 \kappa^2 \eta {\partial \ln T_{\rm c} \over \ln n} \!\right )^2 \sqrt{\tilde \nu\!+\!\frac 1 2}{\xi_{\rm BCS}^2\over \epsilon_0 l} 
\ee
with Bohr radius $a_B$ and the Sommerfeld constant $\alpha_0$.

The contribution due to the external bias consists now of two contributions, 
\be
{1\over C_{\rm ex}}={1\over C_{\rm surf}}+{1\over C_{\rm ind}}
\label{49}
\ee
according to the thick sample limit of surface energy, (\ref{ss}) and (\ref{wallp}), and the induced contribution due to diamagnetic currents, (\ref{walld}). In the following we will discuss them separately.

\subsection{Limit of thick samples}

\begin{figure}
\psfig{file=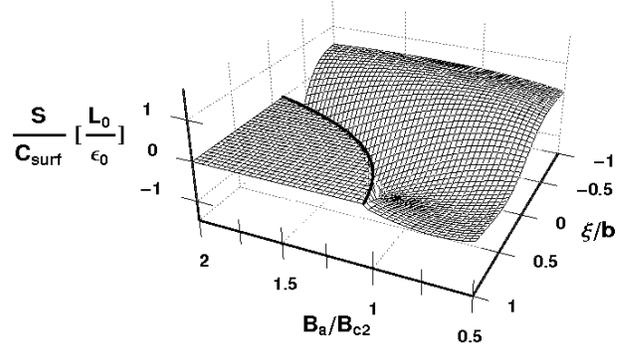,width=8cm}
\caption{The effective change of inverse capacitance in terms of (\ref{L0}) versus the magnetic field and the external bias $1/b$ in units of the coherence length $\xi$. The transition line is plotted as in figure~\ref{surfind}.}
\label{width}
\end{figure}

With the help of the surface energy shown in figure~\ref{surf} we can calculate the surface part of the inverse capacitance (\ref{49}) which is seen in figure~\ref{width}. The larger the magnetic field the larger is the inverse capacitance. For orientation we have plotted the transition line of the minimal eigenvalue of the GL equation where the surface energy is vanishing. 

In order to provide easy to use formulas we can fit in terms of (\ref{L0}) to obtain
\be
\left . {S\over C_{\rm surf}}\right |_0 ={L_0\over \epsilon_0} \left ({l\over \xi}\right )^{3.25} h\left({\xi\over b}\right)
\label{fit1}
\ee
with
\be
h\left(x\right)=1.72 ( x-0.39) {\rm e}^{-2.20 (x-0.39)^2}.
\ee

\begin{table}[h]
\begin{tabular}{|c|c|c|c|c|c|c|c|}
\hline
&$\epsilon_c/n$ & $\kappa_0$& n & $\partial \ln T_c\over\partial \ln n$
& $\partial \ln \gamma\over\partial \ln n$ &$1/\varphi_{\rm el}$ &$L_0/\epsilon_0$
\cr
&[$\mu$eV]&&[$10^{28}$m$^{-3}$]&&& 1/MV& nm$^2$/pF
\cr
\hline
Nb& $4.585$ &0.78&2.2& 0.74\cite{LKMBY07} &0.42\cite{LKMBY07}&4.52&0.248
\cr
YBCO & 750 & 55&0.5&-4.82\cite{note2} &-4.13\cite{note2}&-207.5&2547
\cr
\hline
\end{tabular}
\caption{Used material parameters}
\label{tab}
\end{table}

Using as an estimate the parameters of table~\ref{tab},  one finds
for pure Nb $L_0/\epsilon_0(t=0)=0.248$nm$^2$/pF while for YBCO one gets  $L_0/\epsilon_0(t=0)=2.547$nm$^2$/fF.  This means that the expected change in the inverse capacitance would be some nm$^2$/pF. This should be compared to the trivial part of the inverse capacitance $L/\epsilon_0$. Measuring the distance $L$ between the capacitor and the superconductor in $mm$ one obtains $L/\epsilon_0=112.9 (L/$mm$)$nm$^2$/pF such that a relative precision of 10$^3$ should be required to resolve experimentally the measured effect.

\subsection{Finite width of samples}

\begin{figure}
\psfig{file=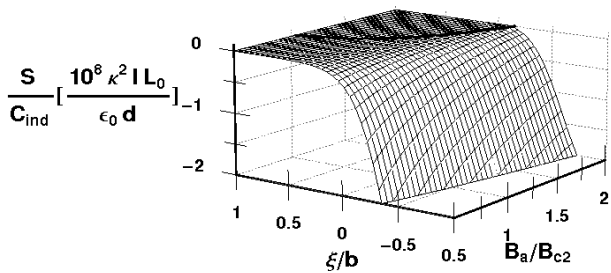,width=8cm}
\caption{The induced part of the effective inverse capacitance in units of the inverse sample thickness scaled with $\kappa^2 \xi$. The transition line is plotted as in figure~\ref{surfind}.}
\label{widthind}
\end{figure}

Now we focus on the induced effect by diamagnetic currents.
The contribution to the effective capacitance (\ref{49}) calculated from (\ref{dll}) is plotted in figure~\ref{widthind}. Compared to figure~\ref{width} one sees a different shape. 
Dependent on the size of the sample the induced effect can be larger than the thick sample limit. 

The induced effect (\ref{walll}) can be  fitted analogously to (\ref{fit1}) as
\be
\left . {S\over C_{\rm ind}}\right |_0= {\kappa^2 l\over d} {L_0\over \epsilon_0} \left ({l\over \xi}\right )^{3.25} h_{\rm ind}\left({\xi\over b}\right)
\ee
with
\be
h_{\rm ind}\left(x\right)&=&-0.028  {\rm e}^{-0.46 (x-0.19)^2}
\nonumber \\
&\times& (0.19+ 0.43 x - 
      1.5 x^2 - 0.16 x^3 + x ^4)
\ee
valid for $\xi/b<1.5$.

It is interesting to observe that both results, the external (\ref{wallp}) as well as the induced one (\ref{dll}) show a nearly quadratic dependence
\be
{C}\sim l^{4}\sim B_a^{-2}
\label{cp}
\ee
on the external magnetic field. 
This qualitative dependence on the magnetic field has been observed in magneto-capacitance measurements.\cite{MHA03} 

\section{Summary and Conclusion}

In this paper we have considered a superconducting layer under the influence of external magnetic fields parallel to the surface and an electric field perpendicular to the surface. We found a nonlinear dependence of the surface critical magnetic field and the surface energy on the magnetic and electric fields. An effective capacitance is introduced which allows to measure these field effects.  The diamagnetic currents induce a magnetic field profile. The selfconsistent equation for the induced magnetization in terms of a superconducting density profile is derived. The induced magnetization represents an important contribution to the surface energy and the effective capacitance. We report an explicit dependence of the surface energy and the effective capacitance on the layer thickness. 
The inversely-linear dependence on the thickness as well as the nearly quadratic dependence on the magnetic field of the inverse capacitance is in qualitative agreement with measurements of magneto-capacitance. We predict a similar behavior to be valid also in capacitor measurements on superconductors. The effective capacitance is found to show a jump at the surface critical field.
An experimental setup of such capacitance measurements is suggested supplemented by  fit formulas.

\acknowledgments

%M. Schreiber is thanked for proof-reading the manuscript.
This work was supported by research plans
MSM 0021620834 and No. AVOZ10100521, by grants
GA\v{C}R 202/07/0597, 202/08/0326 and GAAV 100100712, by PPP project of 
DAAD, by  
DFG Priority Program 1157 via GE1202/06 and the BMBF and 
by European ESF program NES.

\appendix

\section{Derivation of modified GL boundary conditions}\label{boundc}

Let us outline the appearance of the modified boundary conditions (\ref{b1}). Therefore we start with the standard GL equation (\ref{e})
\be
{1\over 2 m}\left(-i\hbar\nabla-e{\bf A}\right)^2\tilde\Psi+
\alpha\tilde\Psi+\beta|\tilde\Psi|^2\tilde\Psi=0
\label{estand}
\ee
supplemented by the standard GL boundary condition that no current flows through the surface,
\be
{\rm Im}\,\left [ {(i \hbar \nabla+e {\bf A}) \tilde\Psi\over\tilde\Psi}\right ]_{x=0}=0.
\label{boundary}
\ee 

The electric field will change the material parameters, i.e. the GL parameters $\alpha$ and $\beta$. Linearizing the GL equation (\ref{estand}) with respect to this induced electric field effect results in $\tilde \Psi=\Psi+\delta \Psi$ where $\Psi$ obeys the GL equation (\ref{estand}) but with different boundary condition. To see this we consider the induced part of the wave function which is written as $\delta \Psi=\Psi \bar \Psi$  where $\bar \Psi$ is proportional to the square of the Thomas-Fermi screening length 
\cite{LCYC96}is very small \cite{LCYC96} such that its explicit influence can be disregarded. However it effects the boundary condition which translates now from (\ref{boundary}) for $\tilde \Psi$ into the form for $\Psi$
\be
{\rm Im}\,\left [ {(i\hbar \nabla-e {\bf A})\Psi\over\Psi}\right |_{x=0}=-\left . {\nabla \bar \Psi\over1+\bar \Psi}\right |_{x=0}.
\ee 
Since  $\bar \Psi\sim E$ the external bias is changing the boundary condition of the GL equation into the form 
\be
{\rm Im}\,\left [ {(i\hbar \nabla-e {\bf A})\Psi\over\Psi}\right |_{x=0}={1\over b_0}+{E\over \varphi_{\rm fe}}\equiv {1\over b}
\label{b1stand}
\ee
in linear order of the external bias. Since the vector potential is real in our case we obtain (\ref{b1}).
The explicit form (\ref{b1}) has been derived \cite{LMKY06} for layered superconductors 
following the DeGennes theory \cite{dG89} or for bulk superconductors.\cite{Mtc01} Please note that even without external bias the term $1/b_0$ exists due to impurities and is usually small.\cite{dG89}

\section{Surface counterterm  in GL equation}\label{paradox}

For simplicity we consider the one-dimensional problem with a superconductor in $x>0$ direction.
The GL free energy density reads
\be
f(x)=f_0+\alpha (x) \Psi(x)^2+\gamma \Psi'(x)^2+\frac \beta 2 \Psi(x)^4
\label{ee}
\ee
where $\alpha(x)=\alpha+{\gamma\over 4 l^2} (\tau_0+x/l)^2 $ and $\gamma=\hbar^2/2 m$. Searching for a minimum of
\be
\delta \int\limits_0^\infty f(x) dx =0
\ee
with the ansatz $\Psi(x)={\cal C} D_\nu(x)$  with respect to the amplitude ${\cal C}$
leads to (\ref{calc}) with (\ref{ab}) without counterterms and (\ref{cc}). This minimum condition can be written also as
\be
0&=&\int\limits_0^\infty \left (\gamma \Psi'^2+\alpha(x) \Psi^2+\beta \Psi^4\right )\nonumber\\
&=&\int\limits_0^\infty\!\! \left [\!-\!\gamma \Psi \Psi''\!+\!\alpha(x) \Psi^2\!+\!\beta \Psi^4\right ]\!-\!\gamma \Psi(0)\Psi'(0).
\label{min}
\ee
This minimum condition differs from the
Lagrange equation of motion as the functional minimization of (\ref{ee})
\be
\partial_x \left ({\delta f\over \delta \Psi'}\right )-{\delta f\over \delta \Psi}=0
\label{lagrange}
\ee
leading to the GL equation
\be
-\gamma \Psi''+\alpha(x)\Psi-\beta \Psi^3=0
\label{GL}
\ee
just by the surface term $\Psi'(0)\Psi(0)$.
For standard GL boundary conditions $\Psi'(0)=0$ this difference does not matter. With the modified GL boundary condition (\ref{b1}), however, we have to add a counterterm to compensate this artifact, i.e.
\be
f(x)\to f(x)+\gamma \Psi'(0)\Psi(0)\delta(x).
\ee

\section{Domain wall surface energy}\label{domain}
Here we outline the derivation of the domain wall parameter (\ref{wallp}) without external bias $b\to\infty$  and for the situation of normal-superconducting boundaries as it is usually found in literature.\cite{Tinkham,FW71,dG89} In terms of (\ref{wavef})
the non-linearized equation (\ref{e}) reads
\be
%&&
-\xi^2 \psi''+{\xi^2 \over 4 l^4 } (x\!+\!\tau_0 l)^2 \psi
%\nonumber\\&&
+\psi-\psi^3=0 
\nonumber\\&&
\label{ee1}
\ee
and
the domain wall parameter (\ref{ss}) for $B_a\approx B_c$ is
\be
\delta(\infty)\!&=&\!\int\limits_0^\infty \!\!dx \!\left [ \psi^4\!+\!2 \xi^2 \psi'^2\!+\! \psi^2\left ({\xi^2\over l^2} (x\!+\!\tau_0 l)^2\!-\!2\right) 
%\right .\nonumber\\ && \left .
 \right ].
\nonumber\\&&
\label{del}
\ee
One can simplify this expression by noting an additional 
conservation law valid for one-dimensional GL equations \cite{LMKMBSa04}.
Multiplying the GL equation (\ref{ee1}) with $\psi(x)$ and integrating over $x$ one gets exactly
\be
\int\limits_{0}^\infty\!\! d x\! \left [
\xi^2 \left ( \!\frac 1 2 (x\!/l+\!\tau_0)^2 \psi^2\!+\!\psi'^2\! \right )\!-\!\psi^2\!+\!\psi^4\right ]=0.
\ee
Subtracting this relation from (\ref{del}) one has the simpler form
\be
\delta(\infty)=\int\limits_{0}^\infty d x \left 
[-\psi(x)^4+\left ({M(x)\over B_c}\right )^2
\right ].
\label{del1}
\ee

The domain wall is characterized by an exponential decay of the magnetic field inside the superconductor and a decay of the superconductor wave function outside. One can calculate the domain wall parameter for two extreme cases analytically.

In strong type-I superconductivity, $\kappa\ll 1$, no magnetic fields are inside, $M(x)=-B_a$. Approximating $\xi^2/l^2\sim B_a\approx 0$ and
with the boundary condition at the surface $\psi(0)=0$ and deep in the superconductor $\psi(\infty)=1$,  the linearized equation (\ref{E1}) has a first integral
\be
\xi^2 \psi'(x)^2=\frac 1 2 (1-\psi(x))^2
\label{integ}
\ee
which can be easily verified by multiplying (\ref{ee1}) with $2 \psi'(x)$ and integrating from $0$ to $\infty$. Taking this first integral (\ref{integ}) into account one finds for (\ref{del}) 
\be 
\delta_{\rm I}(\infty)&=&\int\limits_0^\infty dx \left [ 2 \xi^2 \psi'^2+(1-\psi^2)^2\right ]
\nonumber\\
&=&2\int\limits_0^\infty dx (1-\psi^2)^2=2 \sqrt{2}\xi \int\limits_0^1 d\psi (1-\psi^2)^2
\nonumber\\
&=&\frac 4 3 \sqrt{2} \xi=1.89 \xi
\label{dela}
\ee
where we have used (\ref{integ}) in the second last line once more. We see that in type-I superconductors the surface energy is proportional to the coherence length.

The other case of extreme type-II superconductor is characterized by $\psi\approx 1$ and an exponentially damped  magnetic field profile with the London penetration depth $\lambda$. Therefore (\ref{del}) becomes
\be
\delta_{\rm II}(\infty)=
 \int\limits_0^\infty dx\left [ \left (1-{\rm e}^{-x/\lambda} \right )^2 -1
\right ]=-\frac 3 2 \lambda
\label{delb}
\ee
and we see that the surface energy is proportional to the London penetration depth and becomes negative indicating instability.
We can combine (\ref{dela}) and (\ref{delb})
\be
\delta\approx -\lambda \left (\frac 3 2-{4 \sqrt{2} \over 3 \kappa} \right )
\label{del0a}
\ee
as an interpolation formula for both type-I and type-II superconductors \cite{Tinkham,FW71,dG89}.

%\section{Magnetization}\label{bint}

\section{Induced magnetic field due to diamagnetic currents}\label{bint}

\subsection{Selfconsistent magnetization}

We are going now to investigate the general form of the magnetization provided a profile of the wave function is given. The total induction becomes $B(x,y)=B_a+\mu_0 M(x,y)$ due to the induced magnetization $\mu_0 M(x,y)$ which is determined by the supercurrent ${\bf j}$. For the sake of completeness we discuss also the $y$ dependence.

The external magnetic field and the magnetization are directed in $z$-direction
\be
(0,0,B_a+\mu_0 M(x,y))={\rm rot}{\bf A}
\ee
from which one sees that $A_z=0$ and the remaining possible dependencies of the vector potential are $A_x(x,y)$ and $A_y(x,y)$ and
\be
{\partial A_y\over \partial x}-{\partial A_x\over \partial y}=B_a+\mu_0 M(x,y).
\label{ap1}
\ee
The current is given by
\be
{\bf j}&=&(\hbar {\bf k} -e {\bf A}) {e\over m} |\psi (x,y)|^2
\nonumber\\&=&
{1\over \mu_0} {\rm rot} {\bf B}=\left ({\partial M\over \partial y},-{\partial M\over \partial x},0\right )
\ee   
from which we get the two equations
\be
{\partial M\over \partial y}&=&-{e^2 \over m} |\Psi(x,y)|^2A_x
\nonumber\\
{\partial M\over \partial x}&=&-{e^2 \over m} |\Psi(x,y)|^2 \left (A_y-{\hbar k\over e}\right ).
\label{ap2}
\ee
Eliminating the vector potential in (\ref{ap1}) with the help of (\ref{ap2})
we arrive at the differential equation for the magnetization
\be
{e^2 \mu_0\over m}M={\partial\over \partial x} \left ( {1\over |\Psi|^2} {\partial\over \partial x} M \right )+{\partial\over \partial y} \left ( {1\over |\Psi|^2} {\partial\over \partial y} M \right )
\ee
as the most general equation determining the magnetization profile. It represents a homogeneous linear differential equation of second order which can be solved for the given geometry and a wave function $\Psi(x,y)$ numerically.

Here we restrict ourselves to a slightly simpler geometry where we consider superconductors of large size in $y$-direction, see figure~\ref{exsetup}. This means that the dependence is only $A_y(x)$ and consequently $M(x)$ which means $A_x=0$ according to (\ref{ap2}) and further
\be
A_y(x)=B_a x+\mu_0 \int\limits_0^x dx' M(x').
\label{ay}
\ee
Now we introduce the dimensionless coordinates 
(\ref{coordinates}) 
and 
\be
|\psi(x)|^2={\alpha\over \beta} N D^2(\tau)
\ee
where the wave function $\psi$ is given by the parabolic cylinder functions $D^2(\tau)=D_{\tilde \nu}^2(\tau)$ of (\ref{wave}) only in the linearized ansatz while 
here $D^2$  stands for the square of the appropriately scaled general wave function. In these coordinates eq. (\ref{ap2}) leads then to the integral equation
\be
&&{\mu_0 M(x)\over B_c}=-{N\over \sqrt{2} \kappa} \int\limits_{\tau_0}^\tau d\tau' \tau' D^2(\tau')
\nonumber\\&&
-
{N B_c\over \sqrt{2} \kappa B_a} \int\limits_{\tau_0}^\tau d\tau' D^2(\tau')\int\limits_{\tau_0}^{\tau'} d\tau''{\mu_0 M(\tau'')\over B_c}.
\label{equation}
\ee
Differentiating (\ref{equation}) twice and replacing one integral in the resulting equation by the expression obtained by differentiating (\ref{equation}) once we arrive at the differential equation 
\be
&&h(\tau) y''(\tau)-h'(\tau) y'(\tau)+a h^2(\tau) y(\tau)=0
\nonumber\\
&& y(\tau_0)=1,\qquad y'(\tau_0)=-a \tau_0 h(\tau_0) 
\label{equ}
\ee
for the magnetization 
\be
y(\tau)&=&{\mu_0 M(\tau)\over B_a}+1
\ee
where we
introduced the abbreviations
\be
h(\tau)&=&D^2(\tau)
\nonumber\\
a&=&{N B_c\over \sqrt{2} \kappa B_a}.
\ee

Please note that the trivial solution $y=0$ would mean total diamagnetism and complete Meissner effect. We see from the initial conditions in (\ref{equ}) that this is ruled out. Instead we have a complicated profile. We can alternatively transform (\ref{equ}) by $z(x)=-y'(x)/(y(x) h(x))$ also into a Riccati equation 

\be
z'(x)-h(x) z^2(x)-a&=&0
\nonumber\\
z(\tau_0)&=&a \tau_0.
\label{eq1}
\ee
%{\bf Please note that the differential equation (\ref{equ}) would be exactly solvable if a factor $1/2$ stands in front of $h'(\tau)$ or alternatively, $h(\tau)^3$ instead of $h(\tau)^2$. I checked for mistake but did not find one.} 

\subsection{Approximate form}

Since our wave function (\ref{wave}) serves as a variational ansatz we can restrict the expansion here to terms up to the order ${\cal C}^4$. Therefore we obtain from (\ref{equation})
\be
\!\!\!\!\!\!\!\!\!\!\!\!\!\!\!\!&&{\mu_0M(x)\over B_c}=-{N\over 2 \kappa} \int\limits_{\tau_0}^\tau d \tau' \tau' D^2_{\tilde \nu}(\tau')
\nonumber\\\!\!\!\!\!\!\!\!\!\!\!\!\!\!\!\!&&
\!\!+{N^2 B_c\over 4 \kappa^2 B_a}\!\!\int\limits_{\tau_0}^\tau \!\!\!d \tau' D^2_{\tilde \nu}(\tau')\!\!\!\int\limits_{\tau_0}^{\tau'}\!\!\! d \tau'' \tau'' (\tau'\!\!\!-\!\tau'') D^2_{\tilde \nu}(\tau'')
\!+\!{\cal O}\!(\!N\!)^3.
\label{Mn}
\ee
This expansion can be alternatively considered as an expansion for large applied fields $B_a/B_c$ and/or large GL parameter $\kappa$.

\section{Integrals}\label{integc}
Here we introduce (\ref{Mn}) into (\ref{gnn}) to obtain to additional integral expressions beyond (\ref{f}) proportional to ${\cal B'} N^2$ and ${\cal D'} N^3$. In order to maintain legibility we introduce the integrals
\be
{\cal H}(\tau)&=& \int\limits_{\tau_0}^{\tau} d\tau' D_{\tilde \nu}^2(\tau')
\nonumber\\
{\cal F}(\tau)&=& \int\limits_{\tau_0}^{\tau} d\tau' \tau' D_{\tilde \nu}^2(\tau')
\nonumber\\
{\cal G}(\tau)&=& \int\limits_{\tau_0}^{\tau} d\tau' \tau'^2 D_{\tilde \nu}^2(\tau')
%\nonumber\\
%{\cal L}(\tau)&=& \int\limits_{\tau_0}^{\tau} d\tau' \tau'^3 D_{\tilde \nu}^2(\tau')
\label{fg}
\ee
and write Gibb's free energy (\ref{gnn}) with the abbreviation 
\be
w={d/l}+\tau_0
\ee 
as
\be
&&G=-l {B_c^2\over 2 \mu_0} 
\left (
2 N {\cal A}-N^2 {\cal B} +{N\over \sqrt{2} \kappa} \int\limits_{\tau_0}^{w} d\tau D^2_{\tilde \nu}(\tau) 
\right .
\nonumber\\
&&\left .
\times
\left \{
{B_c N^2\over 2 B_a\kappa^2} \left [\int\limits_{\tau_0}^{\tau} d\tau' {\cal F}(\tau')\right ]^2
+2 \tau \int\limits_{\tau_0}^{\tau} d\tau' 
\left [
-{N\over \sqrt{2} \kappa} {\cal F}(\tau')
\right . \right .
\right .
\nonumber\\&& \left . \left . \left .
+{N^2 B_c\over 2 B_a \kappa^2}
\int\limits_{\tau_0}^{\tau'} d\tau'' {\cal F}(\tau'') ({\cal H}(\tau')-{\cal H}(\tau''))
\right ] 
\right \}
\right .
\nonumber\\
&&
\left .
-\int\limits_{\tau_0}^{w} d\tau 
\left [
{N^2\over 2 \kappa^2} {\cal F}^2(\tau)
-{N^3B_c\over 2 \sqrt{2} \kappa^3 B_a} {\cal F}(\tau) 
\right .
\right .
\nonumber\\&&
\left . \left .\times \int\limits_{\tau_0}^{\tau} d\tau' {\cal F}(\tau') ({\cal H}(\tau)-{\cal H}(\tau'))
\right ]
\right )
\nonumber\\
&& \equiv -l {B_c^2\over 2 \mu_0}
\left (2 N {\cal A} -N^2 ({\cal B}+{\cal B}')+{\cal D}' N^3 \right ). 
\label{gnn1}
\ee

In the following the procedure  will consist in rewriting the integrals into expressions which converge for $w\to \infty$ separating the terms explicitly dependent of $w$. This is achieved by systematically transforming the multidimensional integrals into integrals containing in each integration a weight of $D_{\tilde \nu}^2$.

The additional term in (\ref{gnn1}) proportional to $N^2$ reads 
\be
2\kappa^2 {\cal B}'=\int\limits_{\tau_0}^{w} d\tau \left [{\cal F}^2(\tau)+2 \tau D^2_{\tilde \nu}(\tau) \int\limits_{\tau_0}^{\tau} d\tau' {\cal F}(\tau')\right ].
\ee
Rewriting the first integral 
\be
{\cal F}^2(\tau)&=&\left [\int\limits_{\tau_0}^{\tau} d\tau' \tau' D^2_{\tilde \nu}(\tau')\right ]^2
\nonumber\\
&=&
2 \int\limits_{\tau_0}^{\tau} d\tau' \tau' D^2_{\tilde \nu}(\tau') \int\limits_{\tau_0}^{\tau'} d\tau'' \tau'' D^2_{\tilde \nu}(\tau'')
\label{t1}
\ee
and interchanging integrations twice according to
\be
\int\limits_{a}^{b} d\tau\int\limits_{a}^{\tau} d\tau'=\int\limits_{a}^{b} d\tau'\int\limits_{\tau'}^{b} d\tau
\label{t2}
\ee
we arrive at
\be
2\kappa^2 {\cal B}'&=&2\!\!\int\limits_{\tau_0}^{w} \!\!d\tau \tau D^2_{\tilde \nu}(\tau) \int\limits_{\tau_0}^{\tau} d\tau' \tau' D^2_{\tilde \nu}(\tau')
\left (\tau\!-\!\tau'\!+\!w\!-\!\tau \right )
\nonumber\\
&=&w  {\cal F}^2+{\cal I}
\label{bs}
\ee
with
\be
{\cal I}&=&-2 \int\limits_{\tau_0}^w d\tau \tau D^2_{\tilde \nu}(\tau) {\cal G}(\tau)
\ee
where we have used (\ref{t1}) once more and 
abbreviate 
\be
{\cal H}&\equiv& {\cal H}\left ( w \right )
\nonumber\\
{\cal F}&\equiv& {\cal F}\left ( w \right )
\nonumber\\
{\cal G}&\equiv& {\cal G}\left ( w \right ).
\label{fvw}
\ee

The form (\ref{bs}) shows that ${\cal B}'$ is proportional to the thickness of the sample since all remaining integrals converge to a finite value with ${\cal O}(\exp{(-w}))$.

Next we calculate the term 
${{\cal D'} (\sqrt{2} \kappa)^3 B_a/ B_c}\equiv \tilde {\cal D}'$ in (\ref{gnn1}) with
\be
&&\tilde {\cal D}'=\int\limits_{\tau_0}^w d\tau D^2_{\tilde \nu}(\tau)
\left \{
\left [\int\limits_{\tau_0}^{\tau} d\tau' {\cal F}(\tau') \right ]^2
\right .
\nonumber\\&&
\left .
+2 \tau \int\limits_{\tau_0}^{\tau} d\tau' \tau' \int\limits_{\tau_0}^{\tau'} d\tau'' {\cal F}(\tau'') 
\left [ {\cal H}(\tau')-{\cal H}(\tau'') \right ]
\right \}
\nonumber\\&&
+2 \int\limits_{\tau_0}^{w} d\tau {\cal F}(\tau) \int\limits_{\tau_0}^{\tau} d\tau'{\cal F}(\tau')  \left [ {\cal H}(\tau)-{\cal H}(\tau') \right ].
\ee
In the first term we apply (\ref{t1}) and interchange twice according to (\ref{t2}) and in the second term we apply (\ref{t2}) once to arrive at
\be
&&\tilde {\cal D}'=2\int\limits_{\tau_0}^w d\tau 
\int\limits_{\tau_0}^{\tau} d\tau' \left \{
{\cal F}(\tau){\cal F}(\tau')\left ({\cal H}(w)
-{\cal H}(\tau)\right )
\right .
\nonumber\\&&
\left .
+{\cal F}(w){\cal F}(\tau')\left [{\cal H}(\tau)-{\cal H}(\tau')\right ] \right \}.
\label{z1}
\ee
Employing (\ref{t2}) it is easy to see that
\be
\int\limits_{\tau_0}^w d\tau 
{\cal H}(\tau)
&=&w {\cal H}(w)- {\cal F}(w) 
\nonumber\\
\int\limits_{\tau_0}^w d\tau 
{\cal F}(\tau) 
&=&w { \cal F}(w)- {\cal G}(w).
\label{t3}
\ee
Applying (\ref{t2}) and (\ref{t3})  to (\ref{z1}) yields
\be
\tilde {\cal D}'&=&{\cal H} (w{\cal F}-{\cal G})^2
-2 w {\cal F}  \int\limits_{\tau_0}^w d\tau {\cal H}(\tau){\cal F}(\tau)
\nonumber\\
&& 
+2 {\cal F} \int\limits_{\tau_0}^w d\tau {\cal H}(\tau)
\left [2 \tau  {\cal F}(\tau)-{\cal G}(\tau)\right ]
\nonumber\\
&& 
-2 \int\limits_{\tau_0}^w d\tau {\cal H}(\tau){\cal F}(\tau)
\left [\tau  {\cal F}(\tau)-{\cal G}(\tau)\right ].
\label{c13}
\ee
Applying (\ref{t2}) the first two integrals can be calculated
\be
\int\limits_{\tau_0}^w d\tau {\cal H}(\tau)
\left [2 \tau  {\cal F}(\tau)-{\cal G}(\tau)\right ]&=&w^2 {\cal F H} -{\cal F G} -w {\cal G H} 
\nonumber\\
+&&\!\!\!\!\!\!\!\!\!\!\!\!
2 \int\limits_{\tau_0}^w d\tau \tau D^2_{\tilde \nu}(\tau){\cal G}(\tau)
\label{z2}
\ee
and
\be
\int\limits_{\tau_0}^w d\tau {\cal H}(\tau)
{\cal F}(\tau)&=&{\cal F}(w {\cal H} -{\cal F }) -{\cal G H}+{\cal F}^2 
\nonumber\\
+&&
\!\!\!\!\!\!\!\!\!\!\!\!2 \int\limits_{\tau_0}^w d\tau D^2_{\tilde \nu}(\tau)\left [{\cal G}(\tau)-\tau {\cal F}(\tau)\right ].
\label{z3}
\ee
Successive application of (\ref{t2}) and (\ref{t3})  yields after some straightforward but tedious steps
\be
\!\!\!\!\!\!\!\!\!\!\!\!&&\!\!\!\!\!\int\limits_{\tau_0}^w \!d\tau {\cal H}(\tau){\cal F}(\tau)
\left [\tau  {\cal F}(\tau)\!-\!{\cal G}(\tau)\right ]
={{\cal H}\over 2}\left [
w {\cal F}\left ({w} {\cal F}\!-\!2 {\cal G}\right ) \!+\!{\cal G }^2
\right ]
\nonumber\\
&&-
\frac 1 2  \int\limits_{\tau_0}^w d\tau D^2_{\tilde \nu}(\tau)\left [{\cal G}(\tau)-\tau {\cal F}(\tau)\right ]^2.
\label{z4}
\ee
Using (\ref{z2})-(\ref{z4}) in (\ref{c13}) we arrive finally at
\be
{\cal D}'&=&{B_c\over (\sqrt{2} \kappa)^3 B_a}  \biggl [ w  {\cal F} {\cal J}-{\cal K} \biggr ]
\label{ds}
\ee
where 
\be
\nonumber\\
{\cal J}&=&{\cal F}^2-2 \int\limits_{\tau_0}^{w} d\tau D^2_{\tilde \nu}(\tau) {\cal G}(\tau)
\nonumber\\
{\cal K}&=&
-\int\limits_{\tau_0}^{w} \!d\tau D^2_{\tilde \nu}(\tau) \left [\tau {\cal F}(\tau) -{\cal G}(\tau)\right ]^2
\nonumber\\&& 
+2 {\cal F} \left [{\cal F G}+{\cal I}\right ].
\label{jj}
\ee
We see that (\ref{ds}) as (\ref{bs}) are linearly proportional to the thickness of the sample.

\bibliography{kmsr,kmsr1,kmsr2,kmsr3,kmsr4,kmsr5,kmsr6,kmsr7,delay2,spin,gdr,refer,sem1,sem2,sem3,micha,genn,solid,deform}

\begin{thebibliography}{58}
\bibitem[{not()}]{note2}
\bibinfo{note}{The pair-able charge per Cu atoms is $-0.4335\,e$. In Fig.~2.16
  Plakida\cite{Plak95} shows results of Cava\cite{CHHBMRKPR90} according to
  which the charge transfer $-0.03\,e$ from chains to planes per Cu site leads
  to a decrease of the critical temperature by 30~K. This corresponds to
  ${\partial\ln T_{\rm c}\over \partial\ln n}=-4.82$. From Fig.~3 of
  [\onlinecite{LMCL93}] we can see that the specific heat coefficient drops at
  the same time from $4.4$~mJ/gK$^2$ to $3.0$~mJ/gK$^2$ which gives
  ${\partial\ln \gamma\over \partial\ln n}=-4.13$. This is only a rough
  estimate since the specific heat data include chains while we need the change
  in plane only.}
\end{thebibliography}

\end{document}